\documentclass[pdflatex,sn-mathphys-num]{sn-jnl}

\usepackage{graphicx}%
\usepackage{multirow}%
\usepackage{amsmath,amssymb,amsfonts}%
\usepackage{amsthm}%
\usepackage{mathrsfs}%
\usepackage[title]{appendix}%
\usepackage{xcolor}%
\usepackage{textcomp}%
\usepackage{manyfoot}%
\usepackage{booktabs}%
\usepackage{algorithm}%
\usepackage{algorithmicx}%
\usepackage{algpseudocode}%
\usepackage{listings}%
\usepackage[export]{adjustbox}
\usepackage{subcaption}
\usepackage{caption}
\usepackage{subcaption}
\usepackage{float}
\usepackage{cleveref}
\usepackage{epstopdf, epsfig}

\theoremstyle{thmstyleone}%
\theoremstyle{thmstyletwo}%

\theoremstyle{thmstylethree}%

\begin{document}

\title[Internal Solitary Wave Generation Using A Jet-Array Wavemaker]{Internal Solitary Wave Generation Using A Jet-Array Wavemaker}


\author*[1]{\fnm{Jen-Ping} \sur{Chu}}\email{jenpingc@usc.edu}

\author[2]{\fnm{Patrick} \sur{Lynett}}\email{lynett@usc.edu}

\author[1]{\fnm{Mitul} \sur{Luhar}}\email{luhar@usc.edu}

\affil*[1]{\orgdiv{Department of Aerospace and Mechanical Engineering}, \orgname{University of Southern California}, \orgaddress{\city{Los Angeles}, \postcode{90010}, \state{CA}, \country{USA}}}

\affil[2]{\orgdiv{Department of Civil Engineering}, \orgname{University of Southern California}, \orgaddress{\city{Los Angeles}, \postcode{90010}, \state{CA}, \country{USA}}}


\abstract{This paper evaluates the experimental generation of internal solitary waves (ISWs) in a miscible two-layer system with a free surface using a jet-array wavemaker (JAW).  Unlike traditional gate-release experiments, the JAW system generates ISWs by forcing a prescribed vertical distribution of mass flux. Experiments examine three different layer-depth ratios, with ISW amplitudes up to the maximum allowed by the extended Korteweg-de Vries (eKdV) solution. 

Phase speeds and wave profiles are captured via planar laser-induced fluorescence and the velocity field is measured synchronously using particle imaging velocimetry.  Measured properties are directly compared with the eKdV predictions. As expected, small- and intermediate-amplitude waves match well with the corresponding eKdV solutions, with errors in amplitude and phase speed below 10\%. For large waves with amplitudes approaching the maximum allowed by the eKdV solution, the phase speed and the velocity profiles resemble the eKdV solution while the wave profiles are distorted following the trough. This can potentially be attributed to Kelvin-Helmholtz instabilities forming at the pycnocline. Larger errors are generally observed when the local Richardson number at the JAW inlet exceeds the threshold for instability.}




\maketitle

\section{Introduction}\label{intro}

Internal solitary waves (ISWs) are observed in stratified ocean environments with amplitudes greater than 100 m and wavelengths spanning several kilometers (\citet{Duda2004}). They can be excited by shear from tidal flows moving in opposite directions or from interactions with underwater topography. The transport and breaking of ISWs trigger vertical mixing of sediments, nutrients, and energy in the deep ocean. Trains of ISWs have been captured by synthetic aperture radar (SAR) since the 1980s (\citet{Helfrich2006}). With the development of in-situ thermistor arrays, the first time series of ISWs were recorded in the early 2000's (\citet{Duda2004}). 

ISWs propagate with permanent profiles as a result of the balance between nonlinear amplitude steepening and dispersive flattening (\citet{Cavaliere2021}). As such, ISWs have traditionally been described with nonlinear and dispersive wave equations, most notably the Korteweg–de Vries (KdV) equation. Among earlier developments, KdV equations for ISWs under infinite depth were proposed by \citet{Benjamin_1966} and \citet{Ono1975}. The canonical model equations are derived from the complete Navier-Stokes equations under the assumption of inviscid, incompressible, and density-stratified conditions. However, the applicability of the KdV model is found to be restricted to small-amplitude waves traveling in a system with a relatively small upper layer, as demonstrated by \citet{Koop_Butler_1981}, \citet{Kao_1985} and \citet{Grue1999}. For large-amplitude waves, the KdV wave speed prediction is too fast and the wavelength is too narrow compared with physical observations (\citet{Grue1999, Kodaira2016, Cavaliere2021}). A more suitable model for describing ISWs of larger amplitudes is the extended Korteweg-de Vries (eKdV) equation, also known as the Gardner equation (\citet{Djordjevic1978, Ostrovsky1989}), wherein the incorporation of a cubic nonlinearity better describes nonlinear broadening effects. Derived in a similar manner, the modified Korteweg-de Vrie (mKdV) solution examines large-amplitude waves in a two-layer system; additional details of the derivation will be given in Appendix \ref{appendix:mKdV}.

Experimental findings in the work of \citet{Michallet1997} suggest that while the mKdV solution can accurately describe large-amplitude waves in an immiscible two-layer system, its applicability to other general scenarios may be limited. In order to address the nonlinear dynamics beyond the scope of the eKdV model, \citet{CHOI_CAMASSA_1999} proposed an unsteady generalization of the steady long-wave model of \citet{Miyata1985}, known as the MCC model. This model assumes the wave amplitude is comparable to the total depth and ISW wavelength is much greater than the total depth. To assess the validity of the MCC model, ISW experiments were performed in both miscible and immiscible systems. \citet{Kodaira2016} has shown that the MCC model with a free surface successfully predicts large-amplitude waves in an immiscible two-layer system. Nevertheless, the eKdV solution is demonstrated to be more accurate in a miscible two-layer system under small- to intermediate-amplitude waves by \citet{Cavaliere2021}. 
 
The generation of ISWs in two-layer systems has been investigated extensively over the past five decades. Early efforts include a displacement-type mechanism, proposed by \citet{Koop_Butler_1981}. Subsequent work by \citet{Segur_Hammack_1982} suggested a plunger mechanism which prevented direct perturbation at the interface. \citet{Kao_1985} introduced the gate-type mechanism, which enabled the successful generation of a single soliton with a minimal trailing wave train. The invention of this gate-type mechanism has enabled the comparison of the existing ISW theories (\citet{MiChallet1998,Kodaira2016}) and the examination of wave-structure interactions (\citet{Chen2007,Cui2019}).

However, the generation of ISWs using a gate-type mechanism has some inherent limitations. First, since the primary control is the step-like volume of fluid behind the gate, there is a degree of empiricism involved in the generation of waves of desired amplitude. In fact, to the best of our knowledge, all comparisons between experimental ISW measurements and theoretical predictions have thus far been made based on the measured ISW amplitude, i.e., without \textit{a priori} amplitude prediction. Second, the extreme aspect ratio of the step-like initial volume behind the gate can cause the ISW to break into multiple wave trains (\citet{Kao_1985}). Third, gate release experiments do not allow for the evaluation of wave-wave interactions within a single release. Fourth, the gate release can generate significant disturbances and mixing, leading to longer lived transient effects.
   
To address these limitations, in this study, we extend the Jet-Array Wavemaker (JAW) concept from \citet{Ko2014} to ISW generation. The JAW system allows for precise time-varying control of volumetric inputs and outputs at different depths. This enables the generation of ISWs with prescribed velocity profiles. Moreover, it is possible to generate multiple waves at desired intervals. This paper serves as proof-of-concept for the generation of ISWs using the JAW system, evaluates best practices for wave generation, and identifies potential sources of uncertainty and error. 

All experiments are conducted in a miscible two-layer system comprising a freshwater layer on top of a salt solution layer. The eKdV solution is employed as the reference function for driving the JAW system. Characteristics of the ISWs are captured by using a synchronized planar laser-induced fluorescence (PLIF) and particle imaging velocimetry (PIV) system.  The measured wave profiles and velocity profiles are compared with eKdV predictions to assess the performance of the JAW on ISW generation. Appendix \ref{appendix:mKdV} shows the results obtained using the mKdV solution to force the JAW system. The observed ISW displacements and velocity profiles show better agreement with theory when using the eKdV solution.

The paper is organized as follows. The eKdV solution for generating the ISWs is described in \S2. The experimental set-up and the measurement systems are described in \S3. Potential sources of error for the jet-array wavemaker are discussed in \S4. Measured wave profiles, phase speed, and velocity profiles are compared against theoretical predictions in \S5. Conclusions are presented in \S6.

\section{Theory}
\label{section:th}

\subsection{Extended Korteweg-de Vries Solution}
Building upon the classical Korteweg-de Vries (KdV) equation, the extended Korteweg-de Vries (eKdV) equation, or Gardner equation, incorporates a cubic nonlinear term in the governing equation that captures the dynamics of large-amplitude ISWs: 
\begin{equation}
  \zeta_t + c_0 \zeta_x + c_1 \zeta \zeta_x + c_2 \zeta \zeta_{xxx} + c_3 \zeta^2 \zeta_x = 0.
  \label{eKdV_gov}
\end{equation}
Here $\zeta (x,t)$ is the isopycnal interfacial displacement in the vertical direction, $c_0$ is the phase speed of a linear long wave, and the coefficients $c_1$ and $c_2$ lead the nonlinear and dispersive terms, respectively, assuming two-layer stratification with no mean flow.  The constants appearing in Equation (\ref{eKdV_gov}) are given by:  
\begin{equation}
  {c_0}^2 = \frac{g h_1 h_2 (\rho_2 - \rho_1)}{\rho_1 h_2 + \rho_2 h_1},
  \label{eKdV_c0}
\end{equation}
\begin{equation}
  {c_1} = \frac{-3 c_0 (\rho_1 {h_2}^2 - \rho_2 {h_1}^2)}{2(\rho_1 h_1 {h_2}^2 + \rho_2 h_2 {h_1}^2)},\:
  {c_2} = \frac{c_0 (\rho_2 h_1 {h_2}^2 + \rho_1 h_2 {h_1}^2)}{6(\rho_1 h_2 + \rho_2 h_1)},
  \label{eKdV_c12}
\end{equation}
and
\begin{equation}
  {c_3} = \frac{3 c_0}{{h_1}^2 {h_2}^2} 
          \left[ \frac{7}{8} \left(  \frac{\rho_1 {h_2}^2 - \rho_2 {h_1}^2}{\rho_1 h_2 + \rho_2 h_1}\right)^2 
          -  \left(  \frac{\rho_1 {h_2}^3 + \rho_2 {h_1}^3}{\rho_1 h_2 + \rho_2 h_1}\right) 
          \right].
  \label{eKdV_c3}
\end{equation}
Here $g$ is the gravitational acceleration, $\rho$ and $h$ are the density and depth of each layer with the subscript 1 (2) indicating the upper (lower) layer. Under the assumption of weak two-layer stratification, $\Delta \rho = (\rho_2 - \rho_1) \ll \rho_1$ (\citet{Helfrich2006}), the pycnocline displacement has a solution of the form
\begin{equation}
  \zeta(x,t) = \frac{a}{b + (1-b) \cosh^2[(x-c t)/\lambda]},
  \label{eKdV_zeta}
\end{equation}
in which $a$ is the amplitude of the ISW, $c$ and $\lambda$ are the characteristic phase speed and wavelength, and $b$ is the wave parameter. These parameters are given by:
\begin{equation}
  c = c_0 + \frac{a}{3} \left(c_1 + c_3 \frac{a}{2}\right),\: 
  \lambda = \sqrt{\frac{12 c_2}{(c_1 + c_3\frac{a}{2})a}},\:
  b = \frac{-c_3 a}{2c_1 + c_3 a}.
  \label{eKdV_c, ld, B}
\end{equation}
The eKdV theory yields a mathematical limitation on the ISW amplitude, reached as the characteristic phase speed turns imaginary:  
\begin{equation}
  a_{lim} = \frac{4 h_1 h_2 (h_1-h_2)}{{h_1}^2 + {h_2}^2 + 6 h_1 h_2}.
  \label{eKdV_alim}
\end{equation}
The horizontal components of the layer-averaged velocities are given as:
\begin{equation}
  \overline{u_i(x,t)} = c \left[\frac{\pm \zeta_i(x,t)}{h_i \pm \zeta_i(x,t)} \right],
  \label{eKdV_u_PL}
\end{equation}
where the subscript $i$ refers to layer 1 or 2. The positive sign corresponds to layer-averaged velocity for the upper layer while the negative sign for the lower layer. 

\section{Experimental Facility and Techniques}
\label{section:exp}
\subsection{Jet-Array Wavemaker}
\label{section:jaw}

\begin{figure}
  \centerline{\includegraphics[width = 0.8\textwidth]{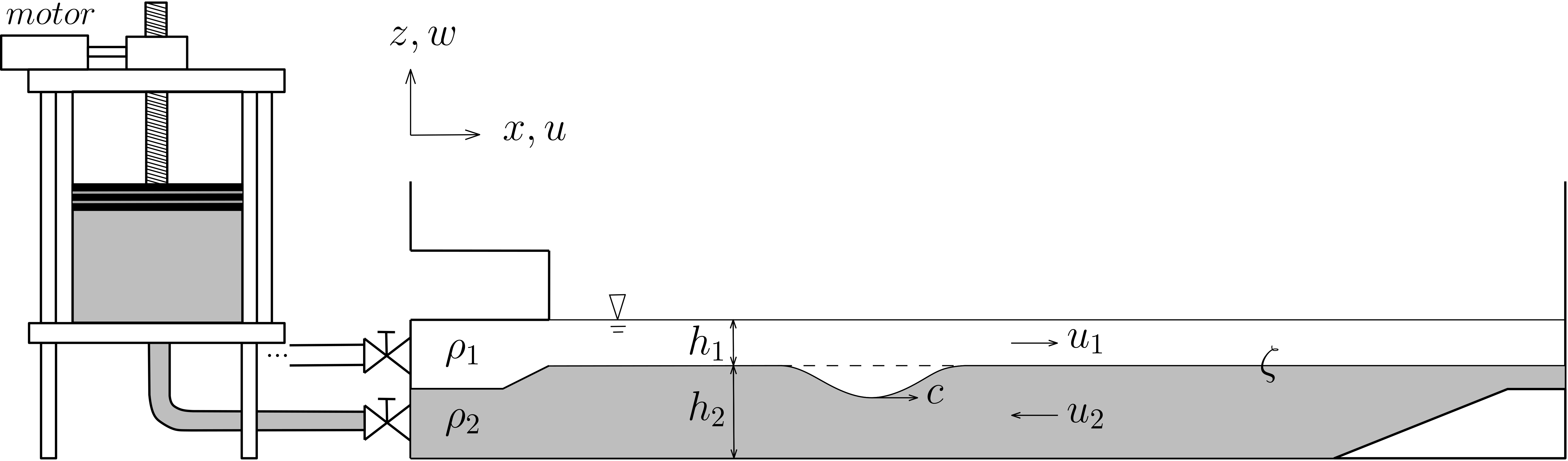}}
  \caption{Experimental setup for internal wave generation in a discrete miscible two-layer system with a free surface. The JAW system is connected to a wave flume.}
  \label{fig:JAW}
\end{figure}

The method to generate ISWs was adapted from the Jet-Array Wavemaker (JAW) configuration used by \citet{Ko2014} to create single-fluid, free surface waves.  The layout of the system is given in Figure \ref{fig:JAW}. Our system comprised two vertically-stacked chambers, or baffles, at the inflow end of a wave flume. Each baffle was connected to an independent fluid reservoir. The dimensions of the cylindrical reservoirs were 12 cm in radius and 25 cm in height.  Each reservoir was capped by a piston, which when moved pushed fluid out or pulled fluid into the reservoir.  The lower channel was connected to the saltwater reservoir, and the upper channel to the freshwater. The flow from the reservoirs, and thus the shape of the ISW, were controlled by the displacements of the pistons, which were driven by Kollmorgen AKM44J servo motors. The servo motors were controlled via a National Instruments USB 6009 data acquisition (DAQ) system capable of multiple digital and analog inputs and outputs. The rotational speed and direction of the motor were controlled by a temporal signal generated via MATLAB in an open-loop control scheme. 

The JAW system was connected to a wave flume of length 2.2 m, width 0.2 m, and total height 0.45 m. The flume was constructed from clear acrylic sheets to allow for optical access.  There was a 1:7 slope with a height of 5 cm at the end of the flume to force ISW breaking and dissipation with minimal reflection. 

The layer volume flux was calculated by multiplying the targeted layer-averaged velocity in Equation (\ref{eKdV_u_PL}) by the cross-sectional area of the baffle. The servo motors moved the pistons to match this volumetric flow rate. This is effectively a simple conservation of mass, control-volume problem. To reduce the turbulence generated at the inlet, two 5 cm-thick honeycomb structures were installed at the inlet in both channels. Downstream of the honeycomb structure, a 5 cm long adjustable ramp was hinged to the boundary separating the lower and the upper layers. The angle of the ramp with respect to the horizontal was adjusted such that the downstream elevation of the ramp matched the layer interface elevation.

\subsection{Two-layer System}
\label{subsection:two-layer}

\begin{table}
\def~{\hphantom{0}}
  \begin{tabular}{lcccccc}
  \toprule
       &  ~$\alpha$ & ~$\beta$ &  ~$\Delta $  &  Miscibility & Mechanism\\[3pt]
       \midrule
       Present Study & -0.091 $\sim$ -0.227 & 0.100$\sim$0.375 & 0.018$\sim$0.021 & miscible & JAW  \\
       \citet{Grue1999} & -0.043 $\sim$ -0.263 & 0.242 & 0.023 & miscible & Gate-type  \\
       \citet{Cavaliere2021} & -0.089 $\sim$ -0.264 & 0.047$\sim$0.207 & 0.027 $\sim$ 0.032 & miscible & Gate-type  \\
       \citet{MiChallet1998} & -0.005 $\sim$ -0.230 & 0.099 & 0.220 & immiscible & Gate-type  \\
       \citet{Kodaira2016} & -0.040 $\sim$ -0.201 & 0.200 & 0.141 & immiscible & Gate-type  \\
       \botrule
  \end{tabular}
  \caption{Experimental parameters in the present and previous ISW studies.}
  \label{tab:comp}
\end{table}

\begin{figure}
  \centerline{\includegraphics[width = 0.8\textwidth]{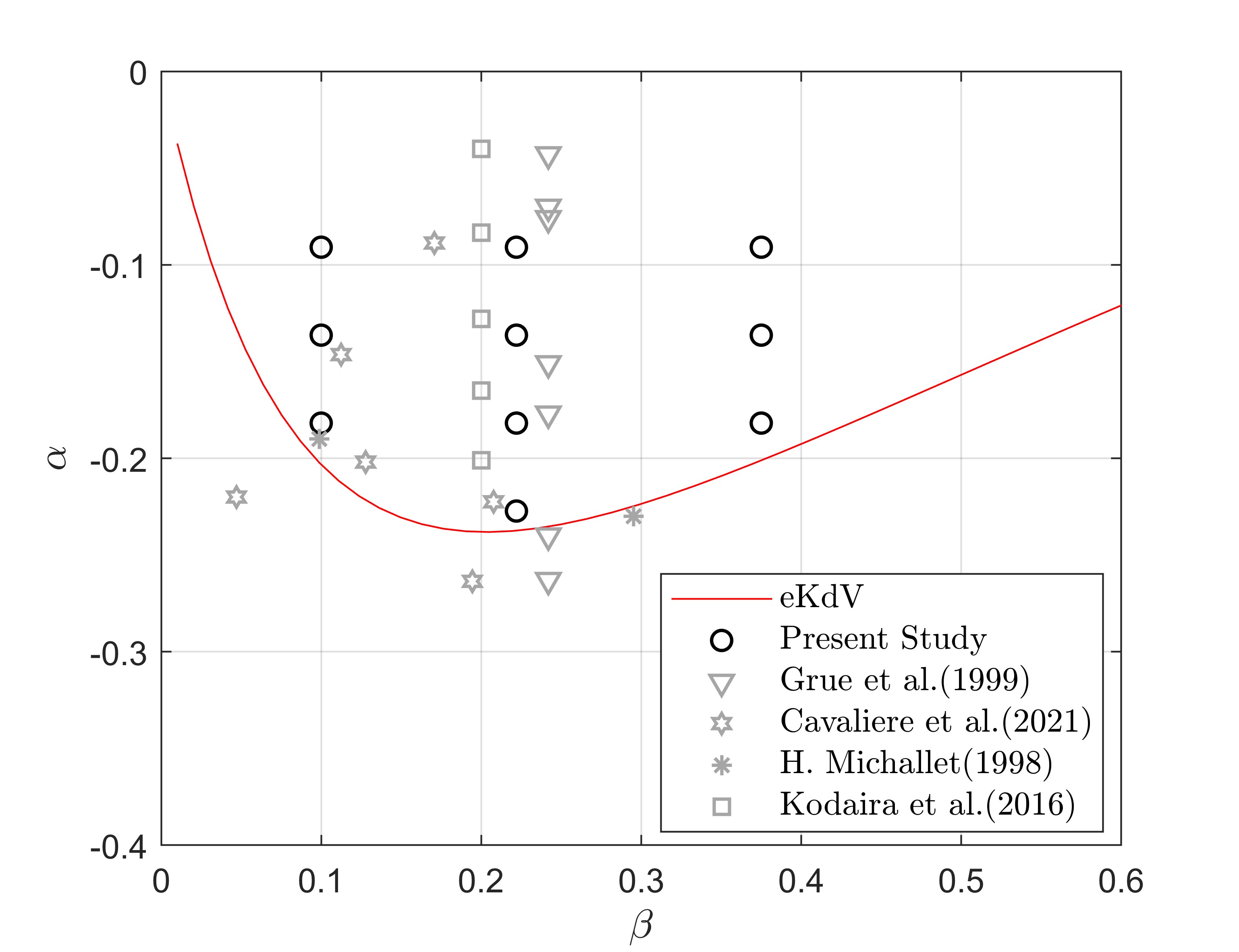}}
  \caption{Dimensionless depth ratio versus dimensionless amplitude for present and prior experimental studies. The red line represents the solution limit of the eKdV model; solutions do not exist below the line.}
  \label{fig:eKdV_comp}
\end{figure}

A miscible two-layer system was set up in the wave flume. The flume was first filled with a salt solution to a depth of $h_2$ with a uniform density of $\rho_2$. The density of the lower layer, $\rho_2$, was set to be 1018$\pm$2 kg/m$^3$. This was confirmed via multiple measurements using a salinity probe. Freshwater with density of $\rho_1 = 998$ kg/m$^3$ ($\rho_1<\rho_2$) was then introduced into the flume through two sponges floating on the free surface of the saline lower layer until the total depth $H = h_1 + h_2$ reached 11 cm. The floating-sponge method minimized mixing between two layers, leading to a stratified two-layer system with a small pycnocline $\Delta \zeta$. 

For comparison purposes, Table \ref{tab:comp} lists the wave parameters and generation mechanisms of ISWs in the present and previous studies.  In this table, $\alpha = a/(h_1+h_2)$ is the dimensionless amplitude relative to the total depth, $\beta = h_1/h_2$ is the dimensionless depth ratio, and $\Delta = (\rho_2-\rho_1)/\rho_1$ is the dimensionless density difference. While most of the earlier studies were performed under a fixed depth ratio to investigate the characteristics of ISWs, this study aims to test the efficacy of the JAW system across a range of conditions. Thus, three nominal depth ratios, $\beta$ = 1/10, 2/9, 3/8, were chosen to examine a broader depth range than found in the previous works. For each depth ratio, wave amplitudes starting from -1 cm ($\alpha = a/(h_1+h_2) \approx -0.091$) up to the amplitude limit predicted by eKdV theory were considered. 

The constraints of the eKdV solution resulted in a smaller maximum $\alpha$ value in this study than in previous research. The $\alpha$ and $\beta$ values presented in Table \ref{tab:comp} are plotted in the parameter map shown in Figure \ref{fig:eKdV_comp}. Black symbols denote the ten experimental runs conducted in the present study. Gray symbols mark all the previous experimental cases utilizing the gate-type mechanism. The red curve outlines the amplitude limitation of eKdV solution, below which no eKdV ISW solution exists. The three black symbols above $\alpha = -0.1$ are referred to as small-amplitude cases, while the three black symbols adjacent to the red curve are identified as large-amplitude cases. The cases positioned in between are classified as intermediate-amplitude cases. As all ISWs in the present study were generated based on eKdV theory, none of the black symbols fall below the red curve.  However, some cases in previous work did exceed the eKdV limitation boundary. Their wave profiles either exhibited significant deviations from theoretical predictions at the trailing edge (\citet{Grue1999, Cavaliere2021}) in miscible systems or were potentially maintained by immiscibility (\citet{MiChallet1998}).

\subsection{Measurement Techniques}
\label{subsection:meas}

\begin{figure}
  \centerline{\includegraphics[width = 0.8\textwidth]{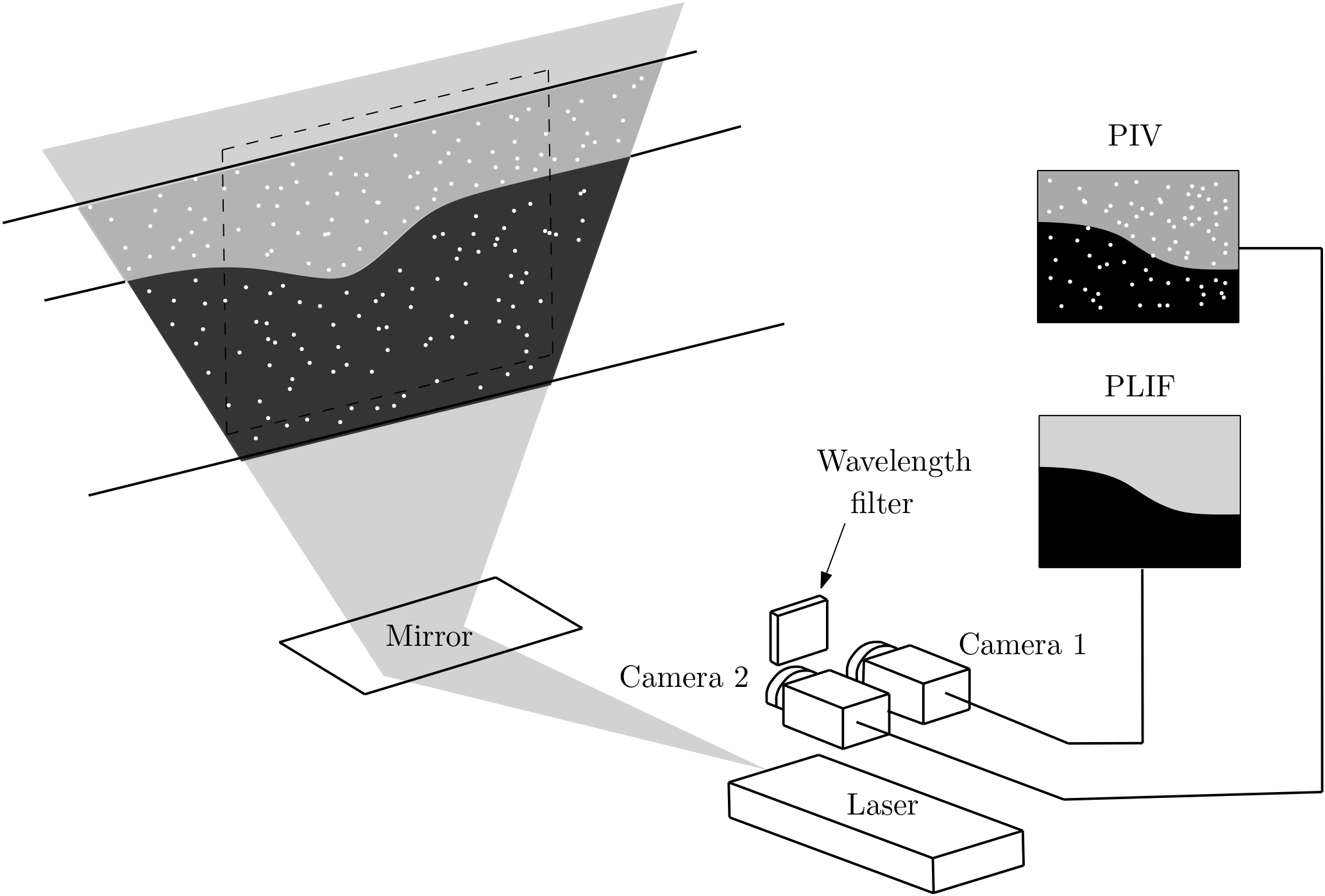}}
  \caption{PLIF, PIV measurement system and data acquisition system}
  \label{fig:Optics}
\end{figure}

The generated ISWs were characterized by synchronizing planar laser-induced fluorescence (PLIF) and two-dimensional two-component particle imaging velocimetry (2D-2C PIV), as shown in Figure \ref{fig:Optics}. The PLIF and 2D-2C PIV systems were composed of a 5W 532nm continuous laser and two CMOS Mako-U-130B 8-bit monochrome cameras with 1024 $\times$ 1280 pixel resolution operated at 30Hz for 40s. The cameras had an overlapping field-of-view of size 16 cm $\times$ 20 cm, positioned at least one wavelength downstream from the inlet.  This location minimized the impact of any initial transients as well as wave reflection from the downstream end of the flume.

\begin{figure}
  \centerline{\includegraphics[width = \textwidth]{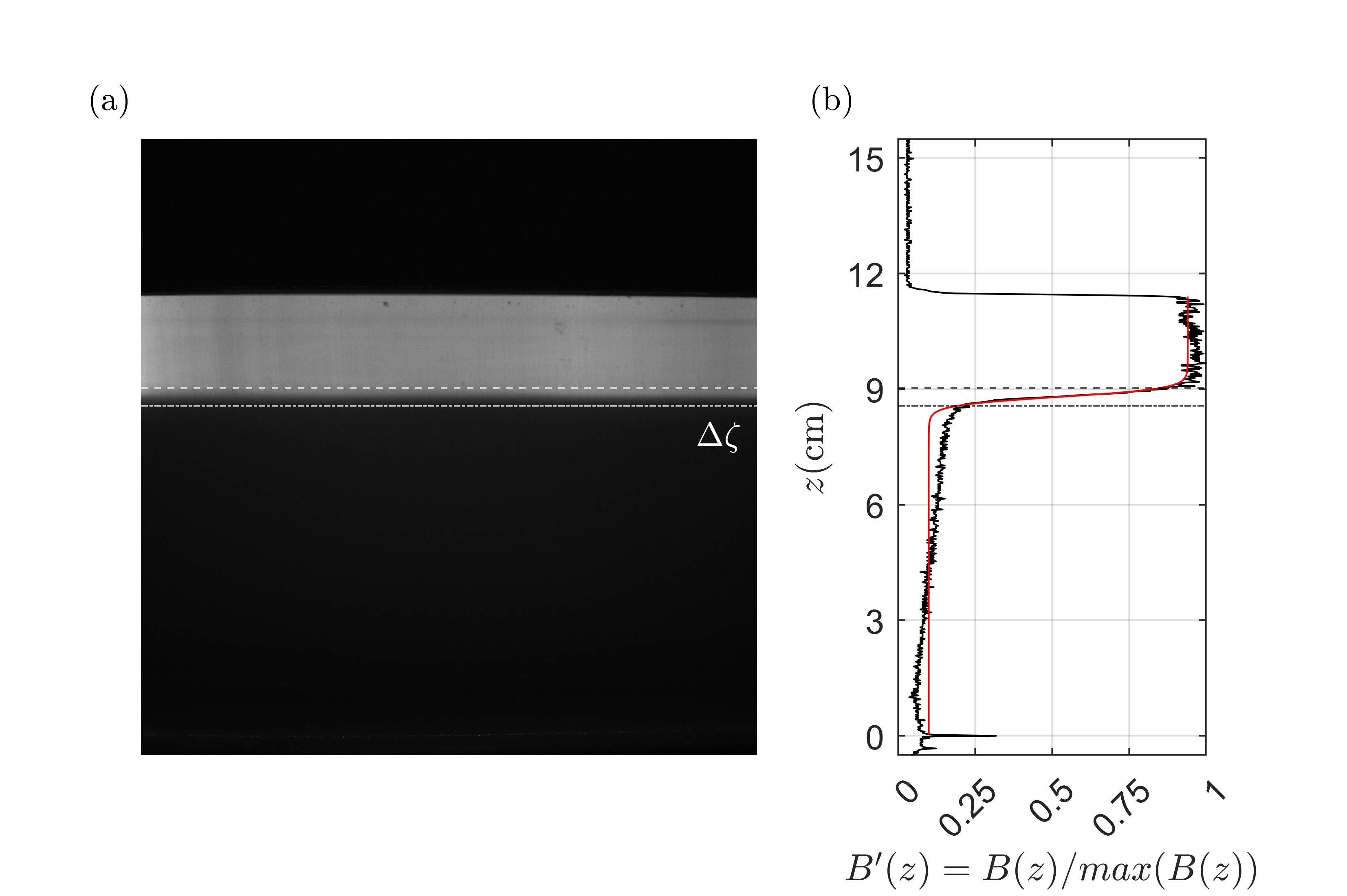}}
  \caption{(a) PLIF visualization of an example initial condition,
         (b) Normalized brightness profile: 
         upper pycnocline(-{}-), lower pycnocline(-.), hyperbolic-tangent approximation (red line)}
  \label{fig:PLIF_IC}
\end{figure}

To separate the upper freshwater layer from the lower layer in PLIF, Rhodamine 6G dye with a peak excitation wavelength of 525nm and a peak emission wavelength of 548nm was premixed with the freshwater to form a homogeneous mixture. The fluorescent region was illuminated vertically by a laser sheet casting from the bottom of the flume. The excited fluorescent region was then imaged using an optical filter with a passband between 536 nm to 564 nm. Assuming single-pixel resolution for the PLIF measurements, the dye field is resolved to approximately 0.16 mm.

Figure \ref{fig:PLIF_IC}(a) shows a snapshot of the initial condition in the flume from a typical configuration. A vertical slice of the brightness profile is normalized by its maximum value as shown in Figure \ref{fig:PLIF_IC}(b). Although the fluorescent fresh layer and the salt solution exhibit homogeneity, the brightness within each layer does not remain uniform. There are slight variations within the upper layer due to obstruction from PIV particles. In the lower layer, light diffusion from above causes a descending brightness trend. 

To estimate the pycnocline thickness, a hyperbolic tangent approximation proposed by \citet{Benney_1966} was fitted to the brightness profile:  
\begin{equation}
  {B^\prime (z)} = B_0 +B_1 \tanh \left[ \frac{2}{\Delta \zeta} (z-\zeta)\right],
  \label{brightness}
\end{equation}
where $B_0$ is the normalized brightness at the pycnocline, $B_1$ is a constant representing the variation, and $\Delta \zeta$ is the pycnocline thickness. To characterize the time-varying wave profile, the vertical position corresponding to $B^\prime (z) = B_0$ was defined as the pycnocline, or interface, across all frames. The upper and lower pycnocline boundaries were assumed to be $\zeta \pm \Delta \zeta/2$. The normalized brightness levels corresponding to the upper and lower bounds of the pycnocline from the initial condition were used to characterize pycnocline thickening. Measured pycnocline thicknesses varied between 0.1 to 0.8 cm (see Table~\ref{tab:exp} below). To characterize the deviation from the nominal depth ratios, layer depths were measured for the initial condition using this brightness profile.  The measured layer depths $h_{1,m}$ and $h_{2,m}$, were defined, respectively, from the free surface to the interface (the location at which $B^\prime = B_0$), and the interface to the bottom of the flume. 

To have PIV particles uniformly distributed across the entire two-layer system, polyamide particles with a diameter of 55 $\mu$m were premixed into the fluorescent fresh water solution. The seeding particles were introduced into the flume at the time of the two-layer system creation. Particle-laden fluid was added through floating sponges placed within and upstream of the region of interest to ensure the highest seeding density for observation before and after ISW propagation. The salt solution was also filtered through a 1 $\mu$m polypropylene cartridge filter to minimize the presence of additional debris. The MATLAB PIVLab package was used to post-process the images using a multi-pass algorithm. The initial pass involved a 64 pixel $\times$ 64 pixel interrogation window.  Two subsequent passes with 32 pixel $\times$ 32pixel interrogation windows with 50\% overlap yielded 69 $\times$ 44 velocity vectors per snapshot. Thus, the PIV measurements have a spatial resolution of approximately 3 mm. Assuming an uncertainty of 1 pixel per frame, the PIV measurements are resolved to roughly 4.8 mm/s. This estimate is conservative since the PIV algorithm resolves displacement fields between image pairs at sub-pixel resolution.

\section{Error Sources}
\label{section: er}

\subsection{Depth Ratio Mismatch}
\label{subsection:DRM}

\begin{figure}
  \centerline{\includegraphics[width = \textwidth]{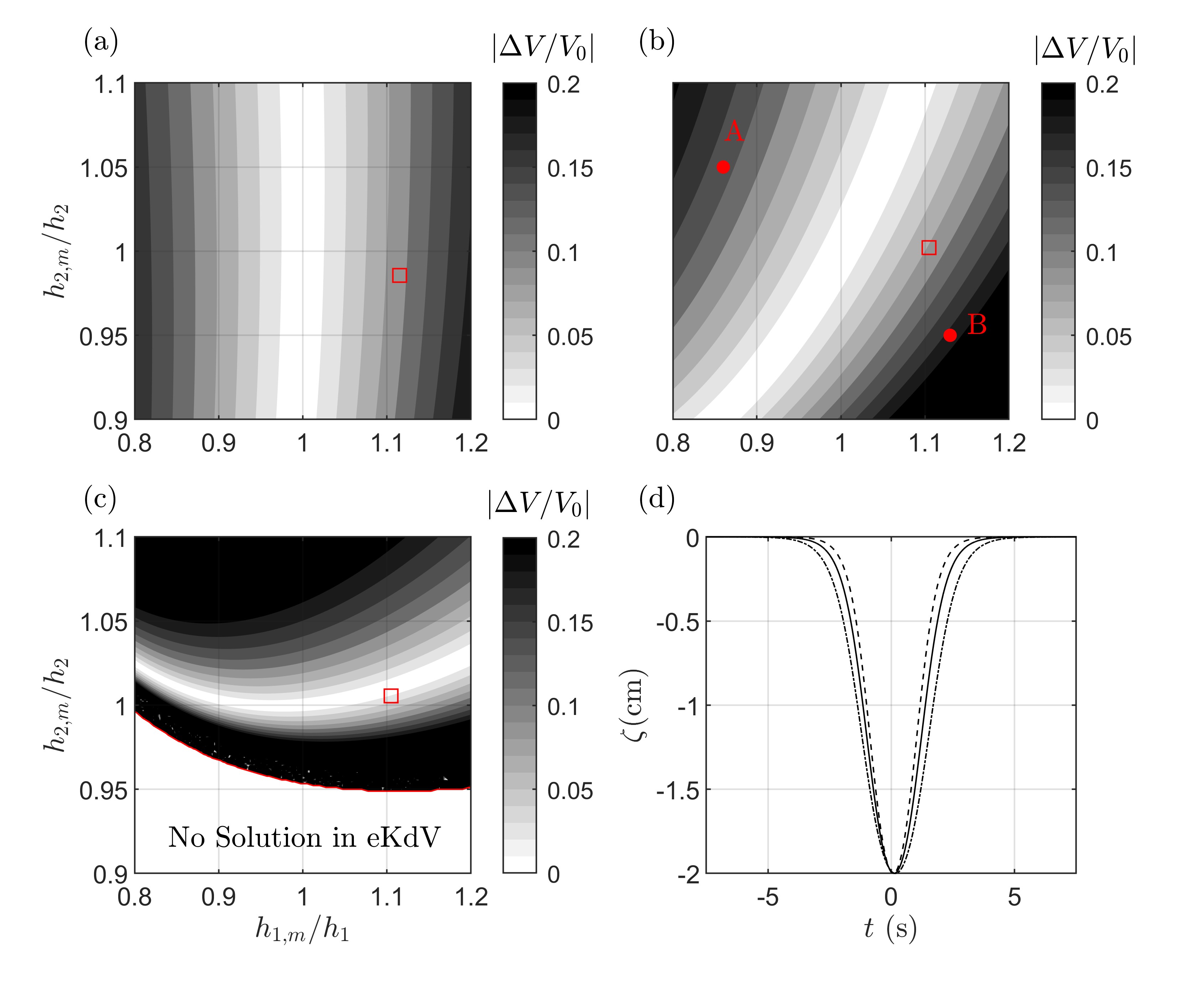}}
  \caption{Sensitivity contours showing volumetric differences due to discrepancies in upper and lower layer depths for nominal depth values $(h_1, h_2) = (2,9)$ cm. The panels show predictions for three different ISW amplitudes: (a) $a = -1$ cm, (b) $a = -2$ cm, (c) $a = -2.5$ cm.  Panel (d) shows wave profiles for $a = -2$ cm. In (a-c), the square markers represent the actual layer depths based on Table \ref{tab:exp}. In (d), the solid line shows wave profiles for the ideal case $(h_{1,m}/h_1,$ $h_{2,m}/h_2) = (1,1)$, the dashed line corresponds to marker A in (b), with $(h_{1,m}/h_1,$ $h_{2,m}/h_2,$ $\Delta V/V_0) = (0.86, 1.05, -0.15)$. The dash-dotted line corresponds to marker B in (b) with $(h_{1,m}/h_1,$ $h_{2,m}/h_2,$ $\Delta V/V_0) = (1.13, 0.94, 0.21)$.}
  \label{fig:Error_sc}
\end{figure}

Creation of the layer depths in our two-layer system has a precision limit. Even with very gradual injection rates, mixing between layers during the filling process and subsequent sponge removal prevented layer depth precision below 3 mm. As a result, the trials had a small mismatch between the actual layer depths and the nominal layer depths. Since the nominal depths were used to drive the JAW and set the elevation of the ramp at the upstream end of the flume, this depth mismatch could lead to wave generation errors. 

The impact of the depth ratio mismatch is characterized in the normalized volumetric difference contours, $\Delta V / V_0$ presented in Figure \ref{fig:Error_sc}. Per the eKdV solution, the velocities in each layer (and hence the fluid volume needed to produce ISWs of desired amplitude) depend on the layer depths. As a result, there is a difference between the volumetric flux used to drive the JAW system, which is calculated from the eKdV solution for the nominal layer depths and desired amplitudes, and the volumetric flux needed for the actual layer depths. This difference is defined as $\Delta V$ and normalized by the nominal volumetric flux $V_0$.   

Panels (a)-(c) in Figure~\ref{fig:Error_sc} show the normalized volumetric difference for the nominal depth ratio $(h_1, h_2) = $ cm for small, intermediate, and large amplitude ISWs. These volumetric differences are plotted as a function of the ratio between measured and nominal layer depths $h_{1,m}/h_1$ and $h_{2,m}/h_2$. The red squares mark the measured values in our experimental work. Brighter regions in the volumetric contours suggest smaller variations in terms of the total volume required to form the wave, while the darker region suggests larger variations.

For the small amplitude case shown in Figure \ref{fig:Error_sc}(a), the vertical alignment of the contours suggests that the volumetric differences are primarily driven by the variations in the upper layer depth; the lower layer has a negligible effect. For the intermediate amplitude case in Figure \ref{fig:Error_sc}(b), variations in both layers have comparable impacts. For the large-amplitude case shown in Figure \ref{fig:Error_sc}(c), the contours are primarily oriented in the horizontal direction, which implies that the volumetric differences are sensitive to the lower layer discrepancies. The blank region in the lower half of Figure \ref{fig:Error_sc}(c) shows the area where no solutions exist according to the eKdV theory. 

Figure \ref{fig:Error_sc}(d) shows wave profiles corresponding to the conditions marked in Figure \ref{fig:Error_sc}(b). The black solid line depicts the profile with no discrepancies in either layer, while the other profiles represent points with volumetric differences larger than $10\%$. Thus, the depth ratio mismatch can lead to either broadening or narrowing of the wave profiles.

For all the experimental results shown in this paper, the mismatch between nominal and actual layer depths resulted in volumetric differences below $10\%$. As a result, a 10\% amplitude error is adopted as a quantitative criterion for successful ISW generation.

\subsection{Wave Generation Errors}
\label{subsection:WGE}

\begin{figure}
  \centerline{\includegraphics[width = 0.65\textwidth]{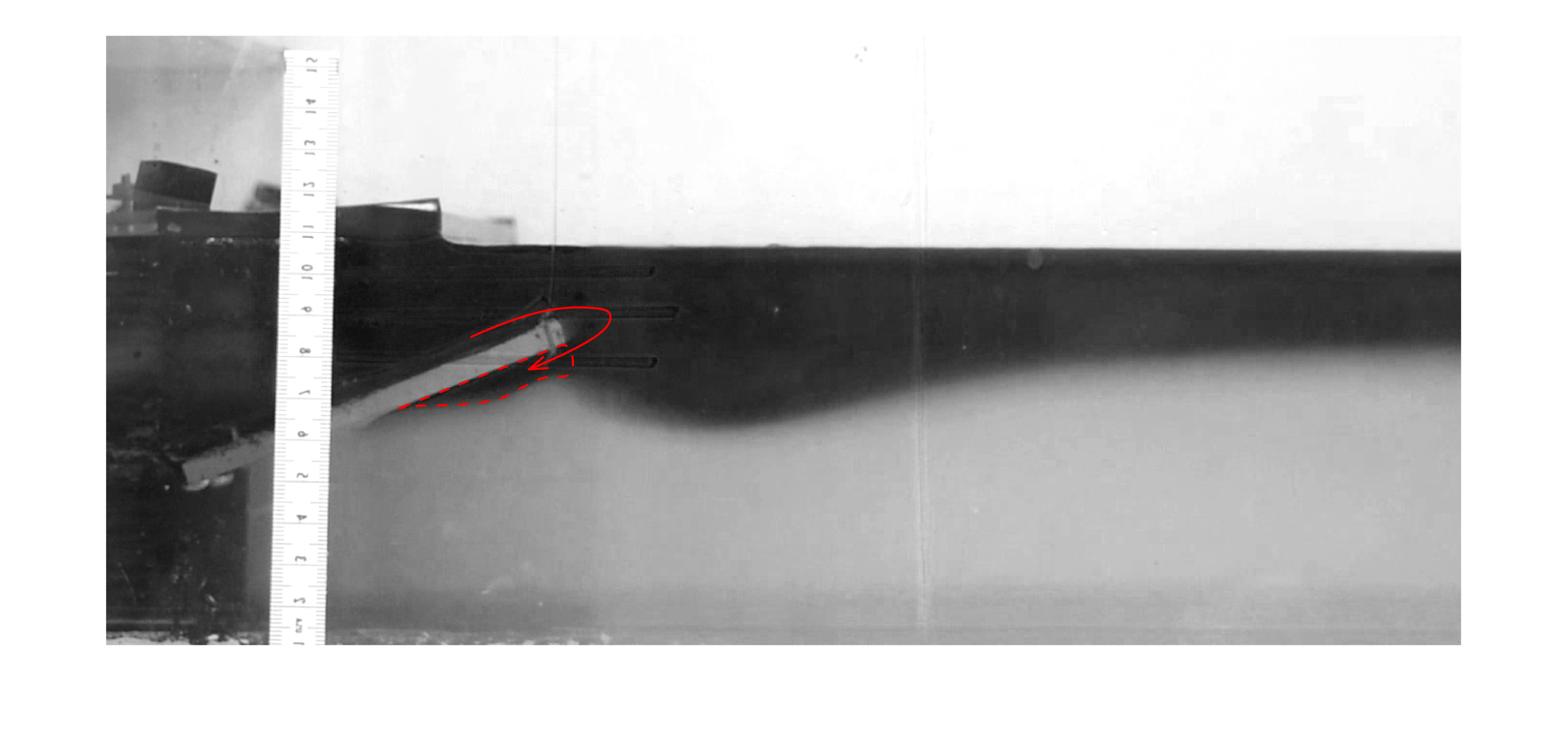}}
  \caption{Dye visualization of large amplitude wave generation at the inlet. The red solid line shows flow direction of the upper layer fluid. The red dashed line shows entrained upper layer fluid below the ramp.}
  \label{fig:KH_en}
\end{figure}

\begin{figure}
  \centerline{\includegraphics[width = \textwidth]{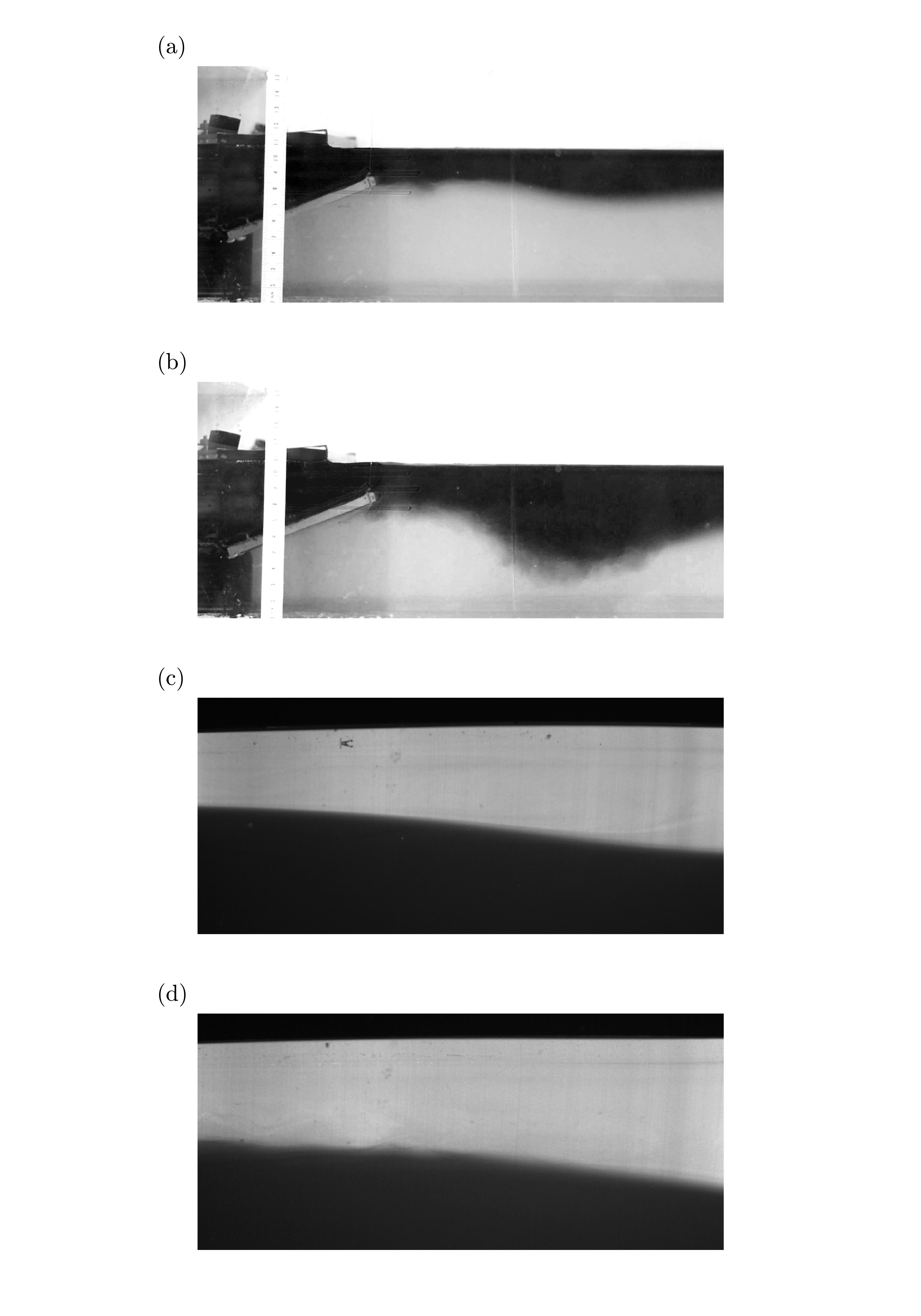}}
  \caption{Dye visualization of small amplitude (a) and large amplitude (b) wave generation at the inlet. PLIF visualizations for case 5 with a sharp pycnocline (c) and case 7 with a diffuse tail due to mixing (d). Wave propagation is from left to right.}
  \label{fig:KH}
\end{figure}

Figure \ref{fig:KH_en} captures the inlet condition as the main depression wave of an ISW departs from the adjustable ramp through dye visualization. For the intermediate and large amplitude cases, there are stronger descending motions of the pycnocline and stronger suction forces from the lower channel. Some upper layer fluid is therefore entrained into the lower layer just below the ramp as shown in Figure \ref{fig:KH_en}. This leads to an upper layer volume deficit in the measured trough of the ISW, which is balanced by an upper layer volume excess on the trailing side of the ISW and in the oscillatory tail (i.e., in any trailing waves). Throughout this manuscript, the term leading side is used to refer to the wave front ahead of the trough (i.e., point of maximum depression). Trailing side is used to refer to the back side of the wave, i.e., after the passage of the trough. 

Figure \ref{fig:KH}(a) shows dye visualization at the inlet for a small amplitude ISW. An interfacial instability can be observed at the end of the adjustable ramp due to the interaction between the jets and the geometry. However, the mixing due to this instability is confined to the inlet region and does not propagate with the ISW. Hence, a sharp pycnocline is observed on the trailing side of the ISW, as illustrated in Figure \ref{fig:KH}(c). 

Figure \ref{fig:KH}(b) shows dye visualization of the inlet condition for a large amplitude ISW. In this case, the larger velocity gradient across the pycnocline leads to a sustained shear driven Kelvin-Helmholtz (KH) type instability at the pycnocline. This KH instability triggers mixing between the lower and upper layers on the trailing side of the wave leading to a thickened pycnocline as shown in Figure \ref{fig:KH}(d). A similar trend was observed in the two-layer system studied by \citet{Kodaira2016}, though the interface remained distinct due to immiscibility.

\section{Results}
\label{section:res}

\subsection{Wave Profiles}
\label{subsection:wp}

\begin{table}
  \centering
  \begin{tabular}{lcccccccccc}
      \toprule
      Case & ~$h_1$(cm)  &  $h_2$(cm) & ~$h_{1,m}$(cm)  &  $h_{2,m}$(cm) & $\rho_2$(kg/m$^3$) & ~$a$(cm) & $a_m$(cm) & $\Delta \zeta_m$(cm) \\[3pt]
      \midrule
       1 & 1.00 & 10.00 & 0.85 & 10.18 & 1017.3 & -1.00 & -0.95  & 0.13\\
       2 & 1.00 & 10.00 & 0.89 & 10.23 & 1017.3 & -1.50 & -1.41  &  0.34\\
       3 & 1.00 & 10.00 & 1.12 & 10.02 & 1017.3 & -2.00 & -1.78  & 0.86\\
       4 & 2.00 &  9.00 & 2.23 &  8.87 & 1019.3 & -1.00 & -0.93  & 0.44\\
       5 & 2.00 &  9.00 & 2.16 &  9.16 & 1019.3 & -1.50 & -1.53  & 0.47\\
       6 & 2.00 &  9.00 & 2.21 &  9.02 & 1019.3 & -2.00 & -2.12  & 0.46\\
       7 & 2.00 &  9.00 & 2.21 &  9.05 & 1016.1 & -2.50 & -2.89  & 0.76\\
       8 & 3.00 &  8.00 & 3.25 &  7.91 & 1018.2 & -1.00 & -1.08  & 0.57\\
       9 & 3.00 &  8.00 & 3.17 &  8.08 & 1020.0 & -1.50 & -1.71  & 0.64\\
       10& 3.00 &  8.00 & 3.18 &  8.11 & 1019.2 & -2.00 & -2.20  & 0.81\\
       \botrule
  \end{tabular}
  \caption{Comparison between measured and nominal values. A subscript $m$ denotes the measured values derived from PLIF.}
  \label{tab:exp}
\end{table}

\begin{figure}
  \centerline{\includegraphics[width = \textwidth]{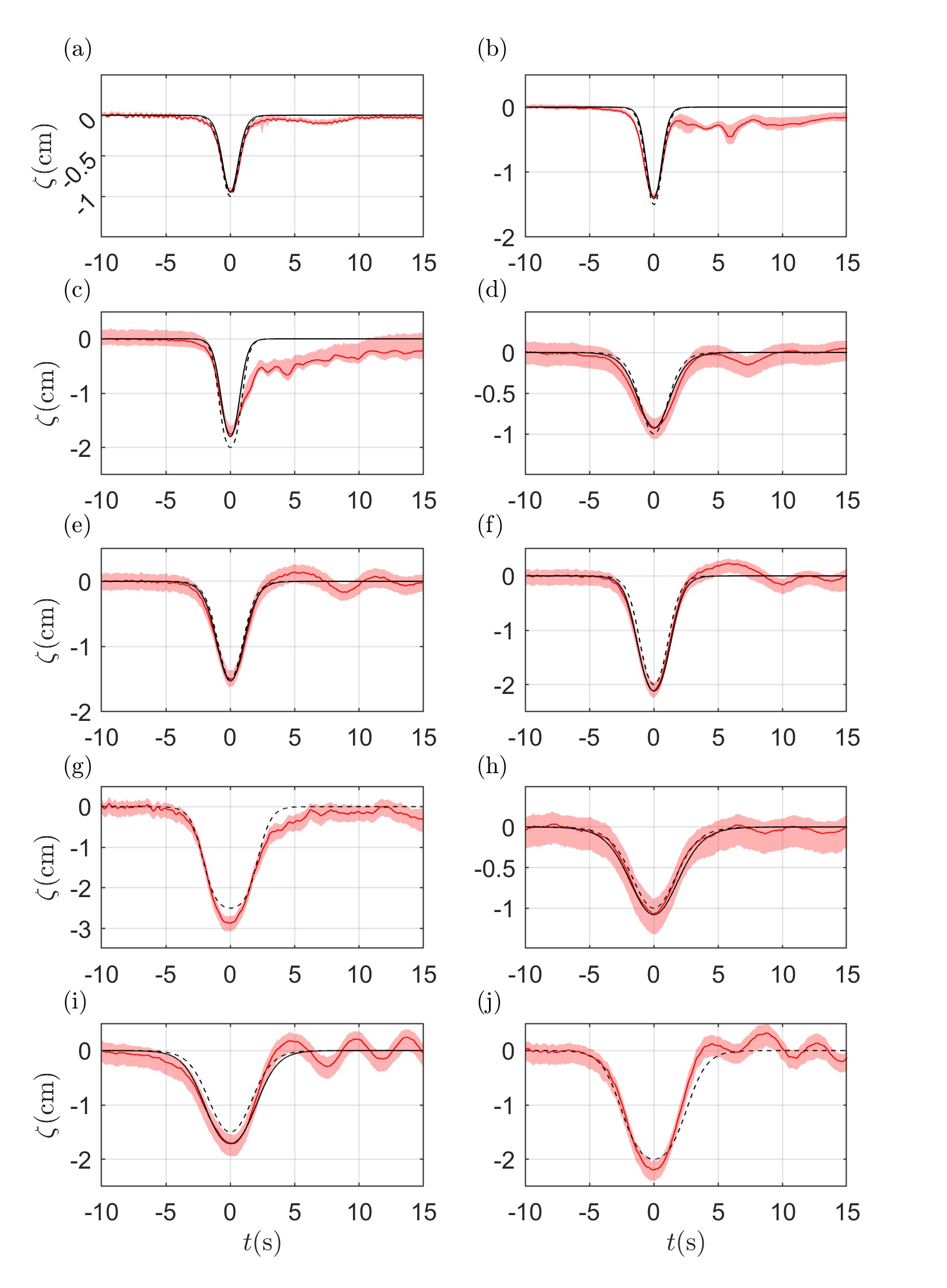}}
  \caption{Comparison between experimental and theoretical wave profiles for nominal depth ratios  $h_1/h_2 = 1/10$ (a-c), $h_1/h_2 = 2/9$ (d-g), and $h_1/h_2 = 3/8$ (h-j). Red lines show the interfacial displacement and red shaded regions show the pycnocline thickness. Black dashed lines show eKdV predictions based on nominal (prescribed) values for depth and amplitude. Black solid lines show eKdV predictions based on measured values.} 
  \label{fig:wp}
\end{figure}

Table \ref{tab:exp} summarizes the parameters for the ten ISW experiments conducted. In this table, $h_1$, $h_2$, and $a$ are the nominal values used to generate the eKdV solution for the JAW, while $h_{1,m}$, $h_{2,m}$ and $a_m$ are the values measured using PLIF. The measured amplitude, $a_m$, is determined by calculating the difference between the mean stationary pycnocline position before wave arrival and the maximum displacement of the pycnocline under the wave trough. Corresponding time series of the wave profiles are provided in Figure \ref{fig:wp}. The shaded red envelopes surrounding the measured profiles mark the pycnocline thicknesses as defined in Section \ref{subsection:meas}. 

The pycnocline thickness at $t = -10$ s characterizes the initial condition for each trial. Subsequent variation in pycnocline thickness captures the mixing dynamics as the ISW propagates. The measured profiles are generally composed of a main depression wave and an oscillatory tail caused by the mismatch between the theory and nature. The wave front appearing before $t=0$ s is termed the leading side. The portion appearing after the passage of the trough at $t=0$ s is termed the trailing side of the wave. 

In all cases, small-scale ($<$ 1mm) disturbances can be observed in Figure \ref{fig:wp} prior to the main depression wave due to small-amplitude and short-wavelength free surface gravity waves created during ISW generation. Similar disturbances can be found in the experiments of \citet{Kodaira2016} and \citet{MiChallet1998}, but caused by the gate-lifting motion and wave reflection from the end wall. In the small and intermediate amplitude cases shown in Figure \ref{fig:wp} (a,d-f,h), the measured wave profiles are closely aligned with the eKdV solutions. The amplitude of the oscillatory tail after the main depression waves is generally within 1 mm. Furthermore, there is negligible temporal variation in pycnocline thickness, suggesting minimal entrainment across the pycnocline. 

However, for large amplitude cases shown in Figure \ref{fig:wp} (c,g,i-j), the measured wave profiles significantly deviate from the dashed black curve predicted based on the nominal values, and instead follow the solid black curve predicted using measured values. This can potentially be attributed to the depth ratio mismatch discussed in Section \ref{subsection:DRM}; for the large amplitude cases, the wave amplitude and volumetric flux are sensitive to minor variations in layer depths. Note that no solid curves are shown in Figure \ref{fig:wp}(g,j) as the measured amplitudes exceed the eKdV solution limits. The measured profiles for these two cases show reasonable agreement with the black dashed curves predicted using nominal values, except in the vicinity of the wave trough where the measured amplitude is greater. This suggests that the eKdV solution captures ISW physics in miscible two-layer systems reasonably well near the amplitude limit. 

The interfacial instability shown in Figure \ref{fig:KH}(b) for large amplitude waves causes the amplitude of the oscillatory tail in Figure \ref{fig:wp} (c,g,i-j) to increase to approximately 5 mm. Additionally, the pycnocline thickening illustrated in Figure \ref{fig:KH}(d) is evident on the trailing side of the ISW. Similar mixing processes starting from the trailing side of the ISW were shown in the miscible two-layer systems of \citet{Grue1999}. 

In Figure \ref{fig:wp} (b,c), the measured wave profiles exhibit a slow return to the initial interface elevation after the wave has passed. The upper layer fluid volume required to form ISWs for the $(h_1, h_2)$ = (1cm, 10cm) depth ratio is relatively small, i.e., the small upper layer depth translates into lower volumetric requirements for the JAW. As a result, any entrainment at the inlet, as shown in Figure \ref{fig:KH_en}, can alter the wave profiles significantly. For the large amplitude case shown in Figure \ref{fig:wp}(c), the effect of this trapped volume couples with KH instability, leading to a significant dispersion on the trailing side of the ISW. This is the only large amplitude case with a measured amplitude under the nominal value.

\subsection{Amplitude and Phase Speed Errors}
\label{subsection:ph_sp}

\begin{table}
\def~{\hphantom{0}}
  \begin{tabular}{lccccccc}
  \toprule
      Case &  ~$\epsilon_a$ (\%) & $\partial u/ \partial z$ (1/s) & $Ri$ & ~$c$ (cm/s)  &  $c_{m}$ (cm/s) & $\epsilon_c$ (\%) \\[3pt]
      \midrule
       1 & -5.00  & 6.83 & 3.06 & 5.6 & 5.1 & -8.9\\
       2 & -6.00  & 9.31 & 0.63 & 6.0 & 5.9 & -1.7\\
       3 & -11.00 & 11.35 & 0.17 & 6.2 & 6.4 & 3.2\\
       4 & -7.00  & 6.19 & 1.22 & 6.8 & 6.8 & 0.0\\
       5 & 2.00   & 8.85 & 0.56 & 7.0 & 6.6 & -5.7\\
       6 & 6.00   & 11.35 & 0.35 & 7.2 & 7.2 & 0.0\\
       7 & 15.60  & 12.65 & 0.21 & 6.7 & 6.9 & 3.0\\
       8 & 8.00   & 5.57 & 1.10 & 7.1 & 6.8 & -4.2\\
       9 & 14.00  & 8.51 & 0.46 & 7.5 & 7.3 & -2.7\\
      10 & 10.00  & 11.13 & 0.21 & 7.5 & 7.6 & 2.7\\
      \botrule
  \end{tabular}
  \caption{Amplitude and phase speed errors and Richardson number estimates. Here, $\epsilon_{a}$ is the amplitude error measured at the ISW trough, $\partial u/ \partial z$ is the nominal shear at the inlet, $Ri$ is the corresponding Richardson number estimate, $c$ is the phase speeds of the ISWs calculated based on the nominal eKdV solution, $c_m$ is the measured phase speed, and $\epsilon_{c}$ is the phase speed error.}
  \label{tab:richardson}
\end{table}

The amplitude error is quantified as the relative difference between the measured and the nominal amplitude values: $\epsilon_a = (a_m - a)/a$. Table~\ref{tab:richardson} shows that for small and intermediate amplitudes, cases 1, 2, 4-6, and 8, amplitude errors are maintained below 10\%.  Higher errors are primarily observed for the large amplitude cases (3, 7, 9, 10). As noted earlier, these cases exhibited a KH type instability startinf from the inlet. 

Since the instability occurs when the destabilizing effect of the interfacial shear outweighs the stabilizing effect of the density gradient at the pycnocline, we can use a Richardson number evaluated near the inlet to indicate its likelihood:  
\begin{equation}
  Ri = \frac{(-g/\rho_2)(\partial \rho /\partial z)}{(\partial u /\partial z)^2} = \frac{(-g/\rho_2)\left[ (\rho_2 - \rho_1)/\Delta \zeta \right]}{\left[ (u_1 - u_2)/d \right]^2}.
  \label{Ri}
\end{equation}
Here $u_1$ and $u_2$ are the theoretical layer-averaged horizontal velocity components calculated from Equation~(\ref{eKdV_u_PL}) based on the nominal values, and $d = 0.5$ cm is the thickness of the inclined ramp separating the two inlet channels. The pycnocline thickness, $\Delta \zeta$, and the layer densities, $\rho_1$ and $\rho_2$, can be obtained from Table \ref{tab:exp}. According to \citet{Grue1999}, the stability criteria for a stratified parallel flow is that the Richardson number should be greater than 0.25.

The small and intermediate amplitude cases with low amplitude errors have inlet Richardson numbers above 0.25 due to relatively small $\partial u/ \partial z $ values, as shown in Table \ref{tab:richardson}. Conversely, for large amplitudes, cases 3, 7, and 10, the Richardson number falls below 0.25 and the amplitude error exceeds 10\%. In case 9, the amplitude error is above 10\% despite its Richardson number above 0.25, suggesting that the depth ratio mismatch could also be a contributing factor. 

The phase speed error is quantified as the relative difference between the measured phase speed and the predicted phase speed calculated from Equation~(\ref{eKdV_c, ld, B}) based on the nominal values: $\epsilon_c = (c_m - c)/c$. The measured phase speed, $c_m$, is determined by measuring trough arrival time between two different locations spaced 12.5 cm apart in the PLIF field of view. For a frame rate of 30 Hz and the spatial resolution of the images, the uncertainty in the measured phase speed is estimated to be roughly 5 mm/s. As shown in Table \ref{tab:richardson}, the phase speed errors remain within the margin of this uncertainty across all amplitude conditions. Notably, positive phase speed errors are observed only in the large amplitude cases when the Richardson number drops below 0.25. The phase speed for the large amplitude ISWs may increase due to additional volume entrainment causing an increase in wave amplitude. Indeed cases 7 and 10 both show an increase in wave amplitude relative to the nominal value and a corresponding increase in wave speed. However, for case 3, the measured wave amplitude is lower than the nominal value yet the phase speed is higher.

\subsection{Velocity Profiles}
\label{subsection:vp}

\begin{figure}
  \centerline{\includegraphics[width = 0.95\textwidth]{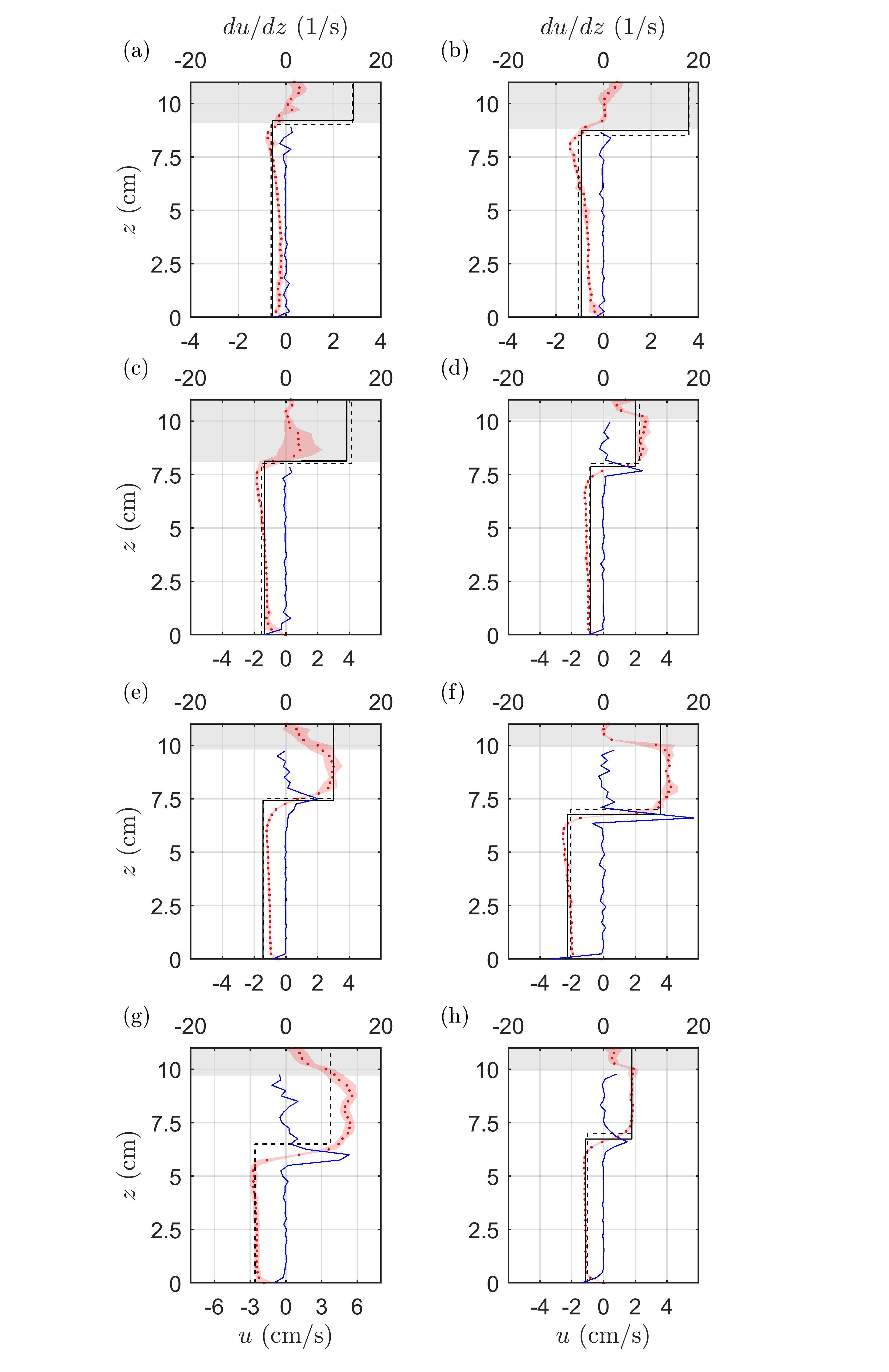}}
  \caption{$continued.$ For description see page ~\pageref{fig:up02}.}
  \label{fig:up01}
\end{figure}

\begin{figure}\ContinuedFloat
  \centerline{\includegraphics[width = 0.95\textwidth]{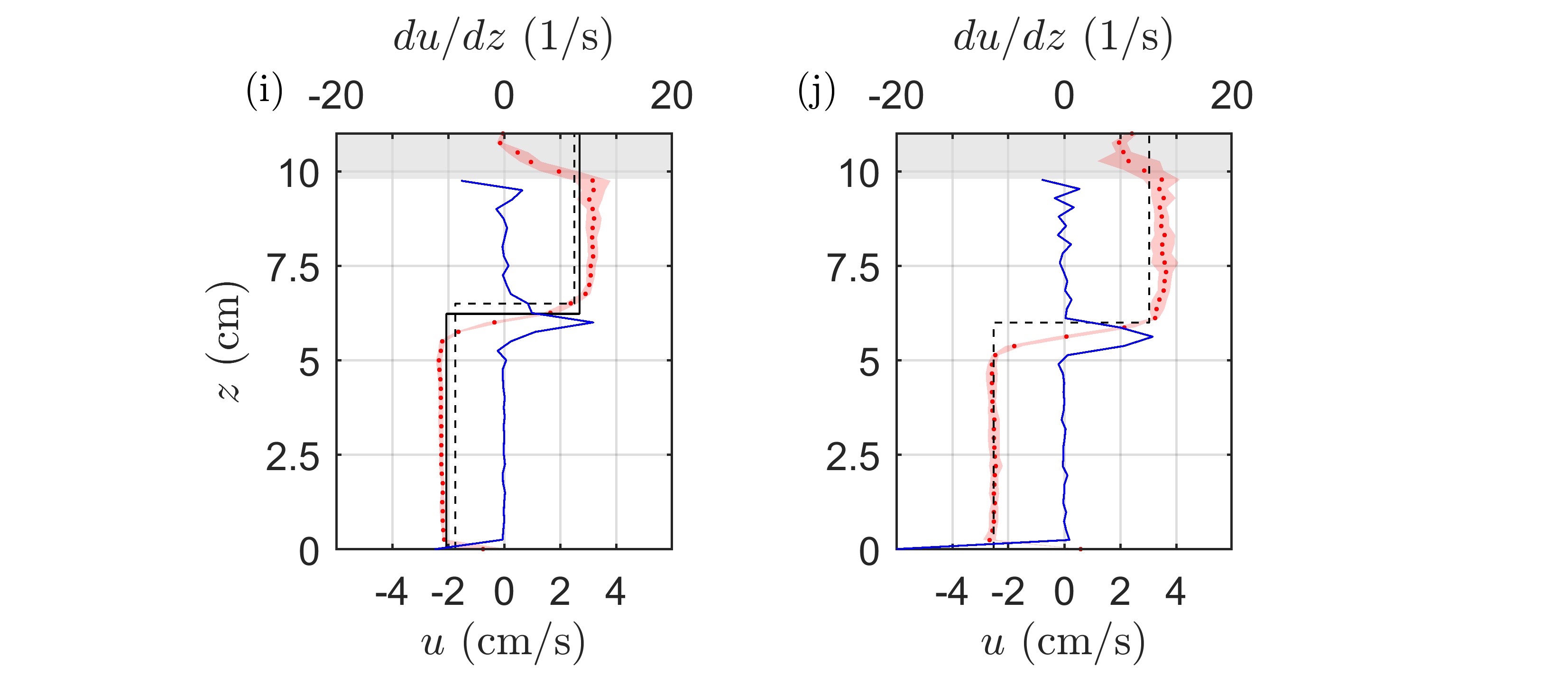}}
  \caption{Comparison between measured and predicted velocity profiles at the wave trough, together with the measured mean shear. The panels correspond to the depth ratios $h_1/h_2 =$ 1cm/10cm (a-c), $h_1/h_2 =$ 2cm/9cm (d-g) and $h_1/h_2 =$ 3cm/8cm (h-j). Red markers and shading show the PIV measurements and one standard deviation. The gray shaded region shows the region without sufficient particles for PIV. Black dashed lines show eKdV predictions based on nominal values. Black solid lines show eKdV predictions based on measured values. Blue sold lines show the shear derived from PIV measurements.}
  \label{fig:up02}
\end{figure}

The horizontal velocity profiles shown in Figure \ref{fig:up01} are extracted from PIV data at the ISW trough. Since the PLIF and PIV are captured synchronously, the PIV-derived velocity profiles are extracted at the time of maximum displacement determined using the PLIF measurement. The red dots in Figure \ref{fig:up01} show the horizontal velocity averaged across a window 6 cm wide in the horizontal direction. The red shaded regions show one standard deviation. 

Despite implementing the floating sponge method to uniformly distribute PIV particles, the uppermost layer remained insufficiently seeded in some of the cases. During the establishment of the two-layer system, the sponges submerged roughly 1 cm as they absorb water, causing the release of particles only beneath a certain depth. These low-seeding regions are shown using gray shading, where the PIV measurements either go to zero or show significantly larger variability. Note that the impact of low seeding is most pronounced for the cases with thin upper layers or for large amplitude waves which lead to a larger downward displacement of the interface. Indeed, for the depth ratio of $h_1/h_2$ = 1cm/10cm, the entire upper layer experiences insufficient seeding. This low seeding region is displaced downwards by the ISWs, yielding unreliable PIV measurements in the entire upper layer as observed in Figure~\ref{fig:up02}(a-c).

As expected, the observed horizontal velocities align with the direction of the wave propagation above the pycnocline and move in the opposite direction below the pycnocline. Moreover, since the lower layer is thicker in all cases, the horizontal velocity in the lower layer is always smaller than that in the upper layer. The measured horizontal velocity profiles are compared with the layer-averaged velocity profiles predicted by the eKdV model, considering both nominal and measured values. 

The measured velocity profiled generally show good agreement with the eKdV predictions in both the lower and upper layers (except for the cases with $h_1=1$ cm where the entire upper layer has insufficient seeding). Differences in predicted velocities using nominal or measured parameters are not significant. 

In all cases, the measured horizontal velocity in the lower layer are nearly uniform. As expected, some of the the profiles show the emergence of a thin viscous boundary layer at the bottom wall. Moreover, the standard deviation in the velocity measurements is small, indicating spatial uniformity.  For the small and intermediate amplitude cases, the high-shear region between the (nearly uniform) upper and lower layers has a vertical extent of $\sim$ 1 cm, as shown in Figure~\ref{fig:up01}(d-f),(h). For the large amplitude cases, however, the vertical transition distance expands to $\sim$1.5-2 cm as illustrated in Figure~\ref{fig:up01}(g),(j). This expansion is consistent with the observed pycnocline thickening, and can potentially be attributed to the mixing induced by KH vortices. Moreover, the upper layer velocities exceed the theoretical prediction by approximately 0.5 cm/s. This is consistent with the observed increase in wave amplitude. However, it must be noted that these small differences in velocity are of the same order of magnitude as the observed variability and the expected PIV uncertainty. Despite the presence of KH instabilities in large amplitude cases, the structure of ISWs in the lower layer remains intact. This is evidenced by the consistent matching of velocity profiles in the lower layer. 

\section{Conclusions}
Internal solitary waves generated by the jet-array wave maker have been experimentally investigated in miscible two-layer systems. Prior work by \citet{Cavaliere2021} suggests that, for a miscible two-layer system, the weakly nonlinear eKdV solution yields good agreement with measurements. Thus, the eKdV solution is selected as the reference for wave generation in the JAW system.

The results presented in this study show that the JAW system can predictably generate ISWs in miscible two-layer systems with varying depth ratios and wave amplitudes up to the eKdV solution limit. Moreover, the generated waves show displacement profiles consistent with theoretical predictions within roughly one wavelength of the inlet. 
Given that minimal mixing is inevitable during the construction process of the two-layer system, discrepancies in depth ratios are identified as potential sources of error. Volumetric sensitivity tests reveal that small amplitude cases are more sensitive to the mismatch in the upper layer depth, while large amplitude cases are more sensitive to the mismatch in the lower layer depth. The absolute volumetric differences were kept below 10\% in all cases to minimize the impact of layer depth mismatch. This effect coupled with pycnocline thickening due to mixing generally led to amplitude errors less than 10\%.

Dye visualization revealed the possibility of wave-generation errors taking place around inlets. For the ISWs with narrow profiles (i.e., the waves with the largest temporal gradients in velocity and pressure at the inlet), the wave can be altered significantly because of flow short-circuiting around the inlet ramp. Due to a suction force in the lower channel, some upper-layer fluid is entrained into the lower layer just below the ramp. In small- and intermediate amplitude cases, instabilities are formed during generation by the interaction between the flow and the geometry in the JAW system. This generational instability is confined near the inlet with minimal influence on the wave formation. Sharpness at the leading side is preserved as the ISW propagates downstream. The measured time series agrees with the analytic solution with minimal oscillation after the main depression wave. 

As the velocity gradient increases with ISW amplitude a sustained KH-type instability is observed at the interface as the wave propagates. KH instabilities overcome the stability provided by density gradients in the large-amplitude cases, triggering mixing across the pycnocline on the trailing side. This additional mixing is thought to be responsible for errors larger than 10\% observed in certain cases. For these cases, larger oscillatory motions left behind by the KH instability and a thickened pycnocline are observed after the main depression wave. Similar instabilities have been documented in prior studies on miscible two-layer ISWs experiments (\citet{Grue1999,Kao_1985}). Local Richardson numbers were estimated around the inlet, where the instability initiates. Increasing amplitude errors and the positive phase speed errors roughly agree with the stability threshold of $Ri = 0.25$ for stratified parallel flows. Despite the varying amplitude errors, the measured phase speed errors are kept within the margin of uncertainty for all cases. 

Through PIV data, we find that the horizontal velocity profiles at the wave trough show strong agreement with the layer-averaged velocity profiles derived from the eKdV solution for small- and intermediate-amplitude cases. Within each layer, the horizontal velocity components remain nearly constant over depth. 

For large-amplitude cases, the shear layer between layers broadens as the pycnocline thickness increases due to the aforementioned KH instability. ISWs with larger amplitude errors induced by KH instabilities result in measured velocities in the upper layer that are at least 0.5 cm/s higher than the eKdV predictions. Conversely, the measured velocities in the lower layer remain well-aligned with the eKdV predictions, suggesting that the deeper lower layers are potentially less impacted by the instabilities.

The strengths and limitations of the JAW system are studied in this work. Previous experimental work has primarily relied on gate release mechanisms to generate ISWs (\citet{Kao_1985, MiChallet1998, Grue1999, Kodaira2016, Cavaliere2021}). 

Yet, it is unclear which of the eKdV, mKdV, or MCC solutions best captures the measurements.  By enabling the generation of ISWs with the theoretically-predicted excitation, the JAW system allows for a direct comparison between theory and experiment. Indeed, the results presented in Appendix \ref{appendix:mKdV} show that the use of the mKdV solution leads to greater discrepancies between the measured profiles and theoretical predictions.

The quality of the ISWs generated by JAW is shown to be comparable with those observed in gate release experiments. Additionally, the JAW system allows for \textit{a priori} amplitude and wave profile selection as well as the generation of multiple solitary waves, periodic internal waves, and surface waves (\citet{Ko2014}). Moving forward, this could enable novel wave-structure (\citet{Michallet1999, Chen2007, Cui2019}) and wave-wave interaction experiments.

\section*{Acknowledgement}
This material is based upon work supported by the National Science Foundation under Grant No. OCE-1830056.  Any opinions, findings, and conclusions or recommendations expressed in this material are those of the authors and do not necessarily reflect the views of the National Science Foundation.

\begin{appendices}
\section{The Internal Solitary Waves in mKdV Solutions}
\label{appendix:mKdV}

The modified Korteweg deVries (mKdV) solution relies on the same governing equations as the eKdV solution of Equation (\ref{eKdV_gov}) but uses the transformation of $\zeta^{'} = \zeta - 2 c_1/c_2$ and the assumption of $c_1 \rightarrow 0$ (\citet{Funakoshi1986, Michallet1997, MiChallet1998, WHOI2006,Cui2019}). This potentially allows for larger density differences, wave amplitudes, and critical depth ratios. The mKdV solution is expressed by
\begin{equation}
  \zeta (x,t) = a \frac{ \mathrm{sech}^2 \left[ (x-ct)/\lambda \right]}{1-\mu \cdot \tanh^2\left[ (x-ct)/\lambda \right]}.
  \label{mKdV_pr}
\end{equation}
The phase speed of the internal solitary waves based on the mKdV solution is 
\begin{equation}
  c = c_0 \left[ 1-\frac{1}{2} \left(\frac{\overline{h} + a}{(h_1+h_2)-h_c} \right)^2 \right],
  \label{mKdV_c}
\end{equation}
where $\overline{h} = h_2-h_c$ marks the distance of the stationary pycnocline position with respect to the critical layer depth, and $c_0$ is the corresponding linear phase speed:
\begin{equation}
  {c_0}^2 = \frac{g(h_1 +h_2)}{2} \left\{ 1- \left[ 1-\frac{h_1 h_2 (\rho_2 - \rho_1)}{\rho_2 (h_1 + h_2)^2} \right]^{1/2} \right\}.
  \label{mKdV_c0}
\end{equation}
The characteristic length of the ISW is
\begin{equation}
  \lambda = 2(h_1 + h_2 - h_c) \sqrt{\frac{(h_1 + h_2 -h_c)^3 + h_c^3}{3(h_1+h_2) h^{'} h^{''}}},
  \label{mKdV_ld}
\end{equation}
with the critical level defined as 
\begin{equation}
  h_c = \frac{h_1 + h_2}{1+\sqrt{\rho_1 / \rho_2}}.
  \label{mKdV_hc}
\end{equation}
The polarity coefficient  
\begin{equation}
    \mu = \begin{cases}
    h^{''}/h^{'} & ,\overline{h} > 0, \\
    h^{''}/h^{'} & ,\overline{h} < 0, 
    \end{cases}
  \label{mKdV_mu}
\end{equation}
is depends on the polarized distances 
\begin{equation}
  h^{'} = -\overline{h} - \left|\overline{h}+a \right|, \quad h^{''} =  -\overline{h} + \left|\overline{h}+a \right|. 
  \label{mKdV_hp}
\end{equation}

\begin{figure}
  \centerline{\includegraphics[width = \textwidth]{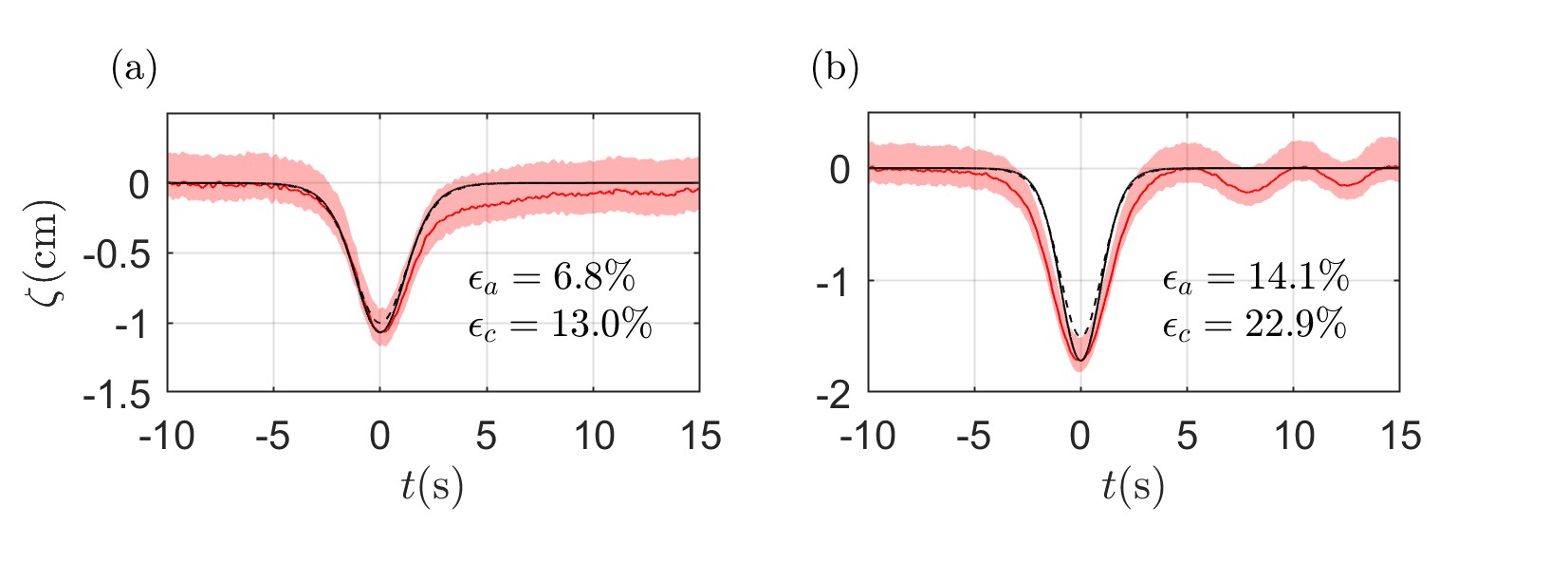}}
  \caption{Comparison between experimental and theoretical wave profiles in time for mKdV profiles at the depth ratio of $h_1/h_2$ = 2cm/9cm for amplitudes $a = -1$ cm (a) and $a = -1.5$ cm (b). Red lines show PLIF interfacial displacement and shaded regions represent pycnocline thickness. Black dashed lines show the mKdV prediction based on nominal values. Black solid lines show the mKdV prediction based on measured values.} 
  
  \label{fig:wp_mKdV}
\end{figure}

\begin{figure}
  \centerline{\includegraphics[width = \textwidth]{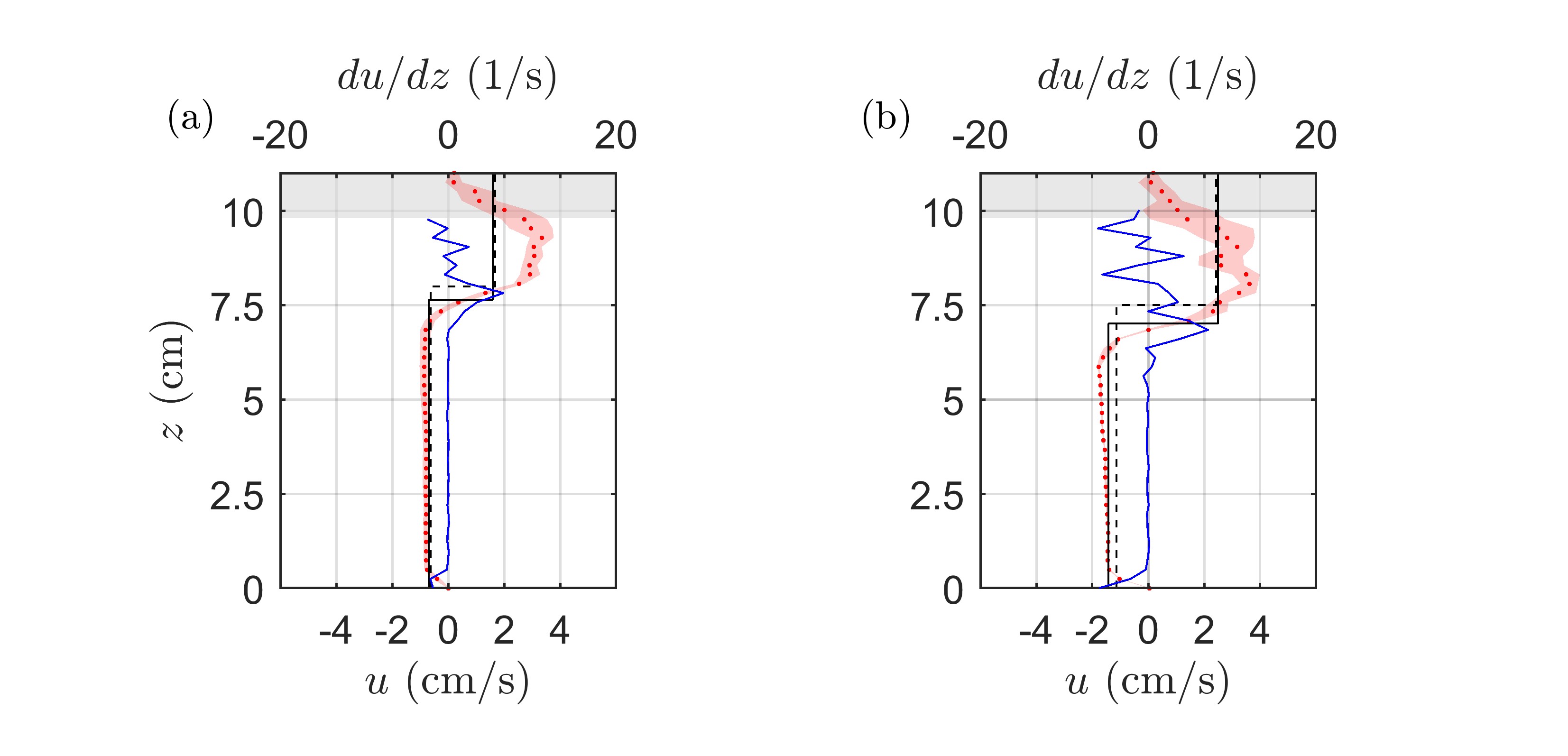}}
  \caption{Comparison between experimental and theoretical velocity profiles at the troughs and the measured shear for mKdV profiles at the depth ratio of $h_1/h_2$ = 2cm/9cm for amplitudes $a = -1$ cm (a) and $a = -1.5$ cm (b). Red shaded regions show measurements within one standard deviation. Gray shading represents regions without sufficient particles for PIV. Black dashed lines show mKdV predictions based on nominal values. Black solid lines show mKdV predictions based on measured values. Blue sold lines show shear derived from PIV measurements.}
  \label{fig:up_mKdV}
\end{figure}

Two cases with the mKdV profiles --- for depth ratio $h_1/h_2 = 2$cm$/9$cm and amplitudes $a = -1$ cm and $a=-1.5$ cm --- are investigated to verify the applicability of mKdV solution in the JAW system (see Section \ref{section:exp}). Because of the assumptions of a negligible $c_1$ term, the dispersion is solely balanced by linear and cubic nonlinearities, resulting in broader wave profiles in the mKdV solution compared to the eKdV solution. In the small-amplitude case as shown in Figure \ref{fig:wp_mKdV}(a), while the amplitude error remains below 10\%, which is comparable to that for the eKdV solutions.  However, the phase speed error exceeds 10\%. Moreover, the excessive volume predicted by the mKdV solution develops into an asymmetric trailing edge after the trough.

As the amplitude increases to $a = 1.5$cm, the larger excessive volume leads to further broadening of the wave profile as observed in Figure \ref{fig:wp_mKdV}(b) along with a 14 \% amplitude growth. The approximately 20\% increase in the phase speed leads to stronger shear, triggering instabilities and oscillatory tail. The wave profiles of the eKdV solution under the same layer depths and amplitudes are presented in Figure \ref{fig:wp}(d, e). Both profiles closely align with the predictions derived from the eKdV solution.

The corresponding horizontal velocity profiles at the troughs are presented in Figure \ref{fig:up_mKdV}. The velocity profiles match the mKdV solution in the lower layer, whereas the excessive volumetric forcing leads to enhanced velocity profiles in the upper layer. The mKdV solution also leads to less uniform velocity profiles in the upper layer during wave propagation compared to the almost-constant profiles shown in Figure \ref{fig:up01} for the eKdV solution.  

As noted in \citet{Michallet1997}, the applicability range of the mKdV solution is limited. The assumption of vanishing $c_1$ term confines its existence only to critical depth ratios of $h_1/h_2 \approx 1$ in miscible two-layer systems or more reasonable depth ratios in immiscible systems with larger density differences. This study examines two preliminary cases of mKdV solution, each sharing identical depth ratios and amplitudes as the eKdV solution detailed in Section \ref{section:res}. Subsequent analysis demonstrates that the eKdV model has a better representation of the ISWs within miscible two-layer systems than does the mKdV model. 



\end{appendices}


\bibliography{sn-bibliography}

\end{document}